\documentclass[10pt]{revtex4}
\usepackage{amssymb}
\usepackage{latexsym}
\usepackage{epsfig}
\usepackage{float}
\usepackage{graphicx,epsfig, color }
\usepackage{graphicx}
\usepackage{subfigure}
\usepackage{amsmath}

\begin{document}
\title{\textbf{Thermodynamics of Static Solutions in (n+1)-dimensional Quintic Quasitopological gravity}}
\author{ A. Bazrafshan$^{1}$, F. Naeimipour$^{1}$, A. R. Olamaei$^{1,2}$ , M. Ghanaatian$^{3}$,}
\address{$^1$ Department of Physics, Jahrom University, 74137-66171 Jahrom, Iran \\
$^2$ School of Particles and Accelerators, Institute for Research in Fundamental Sciences (IPM), P. O. Box 19395-5531, Tehran, Iran, \\
$^3$ Department of Physics, Payame Noor University (PNU), P.O. Box 19395-3697 Tehran, Iran}

\begin{abstract}
Based on the fact that some important theories like string and M-theories predict spacetime with higher dimensions, so, in this paper, we aim to construct a theory of quintic quasitopological gravity in higher dimensions ($n\geq5$). This $(n+1)$-dimensional quintic quasitopological gravity can also lead to the most second-order linearized field equations in the spherically symmetric spacetimes. These equations can not be solved exactly and so, we obtain a new class of $(n+1)$-dimensional static solutions with numeric methods. For large values of mass parameter $m$, these solutions yield to black holes with two horizons in AdS and flat spacetimes. For dS solutions, there are two values, $m_{\rm ext}$ and $m_{\rm cri}$, which yield to a black hole with three horizons for $m_{\rm ext}<m<m_{\rm cri}$. We also calculate thermodynamic quantities for this black hole such as entropy and temperature and check the first law of thermodynamics. Finally, we analyze thermal stability of the $(n+1)$-dimensional static black hole at the horizon $r_{+}$. Unlike dS solutions, AdS ones have thermal stability for each values of $k$, but flat solutions are stable with just $k=1$.  
\end{abstract}



\maketitle

\section{Introduction}
In AdS/CFT correspondence, a duality between strongly coupled conformal field theory and Anti-de sitter gravity is established. According to this correspondence, one can do gravity calculations
to get information about the field theory side or vice versa \cite{Mal}. For example, this correspondence can cause a one-to-one relationship between central charges on quantum side and the coupling parameters on the gravitational side. Einstein gravity is a candidate for this purpose, but, it is not a complete theory and restricts the dual theory to the limited
class of CFT with equal central charges. To remove this limitation, higher-order curvature gravitational theories were introduced in order to provide enough new coupling constants that lead to more central charges in CFT theory. Quasi-topological gravity is a kind of higher-order curvature interaction that depends on its order, one can introduce additional coupling parameters. This gravity has some priorities to other higher curvature theories such as Love-Lock gravity. In Love-Lock theory with terms of order $\mathcal{R}^k$, there is a limitation on dimensions and this order only contributes to the equations of motion with $k\leq[\frac{n}{2}]$, while, in quasitopological gravity with order $k$, there is not such a restriction on dimensions \cite{Deh1}. Quasitopological gravity has also the ability to create linearized equations of motion that coincide with the linearized Einstein's equations up to an overall factor. Not only this linearity can lead us to solve the equations easily, but it also has two other advantages. First, it is possible to find stable vacua in the theory that
are free of ghosts without the breakdown of unitarity in the quantum theory \cite{Sisman}. Second, since the graviton propagator here takes the same structure as the one in Einstein's gravity, holographic studies of the theory would significantly simplified \cite{Myers1,Myers2}.\\
Different orders of quasitopological gravity have been studied so far. Second order of this gravity is matched with the second order of Love-Lock theory (Gauss-Bonnet) and has been investigated in many works \cite{Deh6, Deh7,Hendi3}. Cubic and quartic quasitopological gravity with $\mathcal{R}^3$ and $\mathcal{R}^4$ terms have been also studied in Refs. \cite{Myers3,Brenna1,Naei2} and \cite{Deh1,Naei1,Naei3,Ghana1,Cher} respectively. Recently, a new toy model for gravity in five dimensions has been introduced as quintic quasitopological gravity space \cite{Cist1}. Although this gravity includes curvature tensor of order $\mathcal{R}^5$, but, it leads to equation of motions which are only second-order in derivatives in spherically symmetric spacetimes. Quintic form of quasitopological gravity has some priorities and advantages to other orders of this gravity. First, the obtained solutions of this gravity in asymptotically AdS spacetimes can be duals for a broad class of CFTs. Second, there are five constraints in this theory that arise from the requirement of positive energy 
fluxes \cite{Brig,Ge,Boer} and specify five coupling
constants. Third, based on our information, cubic form of quasitopological gravity is unique, because it has only two possible cubic complete contractions of Weyl tensors, $Tr_{(1)}(C^3)$ and $Tr_{(2)}(C^3)$. Now, not only quintic order is unique against to the quartic form, but also, it may require
at least to properly classify all the non-trivial independent traces of the form $Tr_{(p)}(C^5)$ \cite{Cist1}. Unlike these advantages, Quintic order of quasitopological gravity has a limitation. Quartic quasitopological gravity is the highest order of curvature for which we can obtain exact solutions for black holes. For higher orders like quintic, we are led to field equations which can not be solved exactly and
numerical calculations are challenging to obtain the solutions.\\ 
Higher dimensions are a necessity in gravitational theory for some reasons. Production of higher dimensional
black holes in future colliders can be a possibility in scenarios involving
large extra dimensions and TeV-scale gravity. Also, some important theories such as string and M-theories predict gravity with more than four dimensions. In fact, the first successful statistical counting of black hole entropy in
string theory was performed for a higher-dimensional black hole \cite{Strom1}, that is the best laboratory for the string theory of microscopic black holes. Moreover, black holes are considered as mathematical objects which their spacetimes are found among the most important Lorentzian Ricci-flat manifolds in all dimensions. At last, the subject of charged rotating black holes in
higher dimensions has been investigated in the framework of supergravity theories and string theory \cite{Cvetic1,Cvetic2,Youm}. Based on theses reasons, in this paper, we aim to promote quintic quasitopological gravity from five dimensions to higher ones. Our results can provide an extended investigation in theoretical physics for higher dimensions.\\
The rest of this paper is outlined as follows: In the following section, we construct the general form of the $(n+1)$-dimensional action in quintic quasitopological gravity and then obtain the field equations. In section \eqref{thermo}, we calculate the thermodynamic quantities and then we analyze the physical structure of the solutions in Sec. \eqref{phys} and also investigate their thermal stability in Sec. \eqref{stability}. In the last section, we present our conclusions and remarks.  
\section{Construction of $(n+1)$-dimensional quintic quasitopological gravity}\label{Field}
In this section, we aim to construct quintic curvature terms of quasitopological gravity in higher dimensions. So, we begin with an $(n+1)$-dimensional action which includes higher curvatures up to fifth order and can produce field equations of the second order on spherically symmetric spacetimes. This theory made some interest in
early discussions of higher curvature corrections to the string theory \cite{Boulware,Zwiebach}. The related $(n+1)$-dimensional action in quasitopological gravity is 
\begin{equation}\label{action1}
I_{G}=\frac{1}{16\pi}\int{d^{n+1}x\sqrt{-g}\big\{-2\Lambda+{\mathcal L}_1+\mu_{2}{\mathcal L}_2+\mu_{3}{\mathcal L}_3+\mu_{4}{\mathcal L}_4+\mu_{5}{\mathcal L}_5\big\}},
\end{equation}
where $\Lambda$ is the cosmological constant and
${\mathcal L}_1=R$, ${\mathcal L}_2=R_{abcd}R^{abcd}-4R_{ab}R^{ab}+R^2$, ${{\mathcal L}_3}$ and ${{\mathcal L}_4}$ stand for the Lagrangians of Einstein-Hilbert, Gauss-Bonnet, cubic and quartic curvature corrections in quasitopological gravity respectively with the definitions 
\begin{eqnarray}
{{\mathcal L}_3}&=&
a_{1}R_{ab}^{cd}R_{cd}^{ef}R_{ef}^{ab}+ a_{2}R_{abcd}R^{abcd}R+a_{3}R_{abcd}R^{abc}{{}_e}R^{de}\nonumber\\
&&+a_{4}R_{abcd}R^{ac}R^{bd}+a_{5}R_a{{}^b}R_b{{}^c}R_{c}{{}^a}+a_{6}R_a{{}^b}R_b{{}^a}R +a_{7}R^3,
\end{eqnarray}
and
\begin{eqnarray}
{\mathcal{L}_4}&=& b_{1}R_{abcd}R^{cdef}R^{hg}{{}_{ef}}R_{hg}{{}^{ab}}+b_{2}R_{abcd}R^{abcd}R_{ef}{{}^{ef}}+b_{3}RR_{ab}R^{ac}R_c{{}^b}+b_{4}(R_{abcd}R^{abcd})^2\nonumber\\
&&+b_{5}R_{ab}R^{ac}R_{cd}R^{db}+b_{6}RR_{abcd}R^{ac}R^{db}+b_{7}R_{abcd}R^{ac}R^{be}R^d{{}_e}+b_{8}R_{abcd}R^{acef}R^b{{}_e}R^d{{}_f}\nonumber\\
&&+b_{9}R_{abcd}R^{ac}R_{ef}R^{bedf}+b_{10}R^4+b_{11}R^2 R_{abcd}R^{abcd}+b_{12}R^2 R_{ab}R^{ab}\nonumber\\
&&+b_{13}R_{abcd}R^{abef}R_{ef}{{}^c{{}_g}}R^{dg}+b_{14}R_{abcd}R^{aecf}R_{gehf}R^{gbhd},
\end{eqnarray}
where the coefficients $a_{i}$'s and $b_{i}$'s are defined in the appendix $\eqref{app}$. According to \cite{Cist1}, there are at most 24 terms to construct a general form of Lagrangian containing $\mathcal{R}^5$ terms in quintic quasitopological gravity as follows
\begin{eqnarray}\label{quintic}
{\mathcal{L}_5}&=&
c_{1} R R_{b}^{a} R_{c}^{b} R_{d}^{c} R_{a}^{d}+c_{2} R R_{b}^{a} R_{a}^{b} R_{ef}^{cd} R_{cd}^{ef}+c_{3} R R_{c}^{a} R_{d}^{b} R_{ef}^{cd} R_{ab}^{ef}+c_{4} R_{b}^{a} R_{a}^{b} R_{d}^{c}  R_{e}^{d} R_{c}^{e}\nonumber\\
&&+c_{5} R_{b}^{a} R_{c}^{b} R_{a}^{c}  R_{fg}^{de} R_{de}^{fg}+c_{6} R_{b}^{a} R_{d}^{b} R_{f}^{c}  R_{ag}^{de} R_{ce}^{fg}+c_{7} R_{b}^{a} R_{d}^{b} R_{f}^{c} R_{cg}^{de} R_{ae}^{fg}+c_{8} R_{b}^{a} R_{c}^{b} R_{ae}^{cd} R_{gh}^{ef} R_{df}^{gh}\nonumber\\
&&+c_{9} R_{b}^{a} R_{c}^{b} R_{ef}^{cd} R_{gh}^{ef} R_{ad}^{gh}+c_{10} R_{b}^{a} R_{c}^{b} R_{eg}^{cd} R_{ah}^{ef} R_{df}^{gh}+c_{11} R_{c}^{a} R_{d}^{b} R_{ab}^{cd} R_{gh}^{ef} R_{ef}^{gh}+c_{12} R_{c}^{a} R_{d}^{b} R_{ae}^{cd} R_{gh}^{ef} R_{bf}^{gh}\nonumber\\
&&+c_{13} R_{c}^{a} R_{d}^{b} R_{ef}^{cd} R_{gh}^{ef} R_{ab}^{gh}+c_{14} R_{c}^{a} R_{d}^{b} R_{eg}^{cd} R_{ah}^{ef} R_{bf}^{gh}+c_{15} R_{c}^{a} R_{e}^{b} R_{af}^{cd} R_{gh}^{ef} R_{bd}^{gh}+c_{16} R_{b}^{a} R_{ad}^{bc} R_{fh}^{de} R_{ci}^{fg} R_{eg}^{hi}\nonumber\\
&&+c_{17} R_{b}^{a} R_{de}^{bc} R_{cf}^{de} R_{hi}^{fg} R_{ag}^{hi}+c_{18} R_{b}^{a} R_{df}^{bc} R_{ac}^{de} R_{hi}^{fg} R_{eg}^{hi}+c_{19} R_{b}^{a} R_{df}^{bc} R_{ah}^{de} R_{ei}^{fg} R_{cg}^{hi}+c_{20} R_{b}^{a} R_{df}^{bc} R_{gh}^{de} R_{ei}^{fg} R_{ac}^{hi}\nonumber\\
&&+c_{21} R_{cd}^{ab} R_{eg}^{cd} R_{ai}^{ef} R_{fj}^{gh}R_{bh}^{ij}+c_{22} R_{ce}^{ab} R_{af}^{cd} R_{gi}^{ef} R_{bj}^{gh}R_{dh}^{ij}+c_{23} R_{ce}^{ab} R_{ag}^{cd} R_{bi}^{ef} R_{fj}^{gh}R_{dh}^{ij}+c_{24} R_{ce}^{ab} R_{fg}^{cd} R_{hi}^{ef} R_{aj}^{gh}R_{bd}^{ij}. 
\end{eqnarray}
To find the coefficients $c_{i}$'s, we use a spherically symmetric spacetime with $(n+1)$-dimensional metric
\begin{eqnarray}\label{metr}
ds^2=-f(r)dt^2+\frac{1}{g(r)}dr^2+r^2 d\Omega^2,
\end{eqnarray}
where $d\Omega^2$ shows the line element of a $(n-1)$-dimensional hypersurface with constant curvature $(n-1)(n-1)k$ and volume $V_{n-1}$. The spatial part of the metric, $d\Omega^2$, is defined as
\begin{equation}
d\Omega^2=\left\{
\begin{array}{ll}
$$d\theta^{2}_{1}+\sum_{i=2}^{n-1}\prod_{j=1}^{i-1} \rm sin^2 \theta_{j} d\theta_{i}^2$$,\quad \quad\quad\quad \quad\quad\quad\quad\quad\quad  \ {k=1,}\quad &  \\ \\
$$\sum_{i=1}^{n-1} d\phi_{i}^{2}$$,\quad\quad\quad\quad\quad\quad \quad\quad\quad\quad\quad\quad\quad\quad\quad\quad\quad\quad  \ {k=0,}\quad &  \\ \\
$$d\theta^{2}_{1}+\rm sinh^2 \theta_1 d\theta_{2}^{2}+\rm sinh^2 \theta_{1} \sum_{i=3}^{n-1}\prod_{j=2}^{i-1} \rm sin^2 \theta_{j} d\theta_{i}^2$$, \quad{k=-1.}\quad &
\end{array}
\right.
\end{equation}
where the parameters $k=-1,0,1$ correspond to hyperbolic, flat and spherical geometries respectively. In the spherically symmetric spacetime \eqref{metr}, the quintic quasitopological Lagrangian \eqref{quintic} yields to a second order field equation in higher dimensions, if we choose the coefficients $c_{i}$ as what we have listed in the appendix \eqref{app}. For $n=4$, these higher dimensional coefficients reduce to the ones in Ref. \cite{Cist1}. \\
If we use the above definitions in action \eqref{action1} and then integrate by part, we obtain the effective action  
\begin{eqnarray}\label{Act2}
S=\frac{(n-1)}{16\pi L^2}\int d^{n} x\int{dr \sqrt{\frac{f}{g}}\bigg\{\bigg[r^n\bigg(\hat{\mu}_{0}+\Psi+\hat{\mu}_{2}\Psi^2+\hat{\mu}_{3}\Psi^3+\hat{\mu}_{4}\Psi^4+\hat{\mu}_{5}\Psi^5\bigg)\bigg]^{'}\bigg\}},
\end{eqnarray}
where $\Psi=\frac{L^2}{r^2}\bigg(k-f\bigg)$ and prime shows the derivation with respect to the radial coordinate $r$. We have also used the dimensionless coefficients 
\begin{eqnarray}
\Lambda=-\frac{n(n-1)\hat{\mu}_{0}}{2L^2}
\,\,\,\,,\,\,\,\mu_{2}=\frac{\hat{\mu}_{2} L^2}{(n-2)(n-3)},
\end{eqnarray}
\begin{eqnarray}
\mu_{3}=\frac{8(2n-1)\hat{\mu}_{3}L^4}{(n-2)(n-5)(3n^2-9n+4)},
\end{eqnarray}
\begin{eqnarray}
\mu_{4}=\frac{\hat{\mu}_{4}L^6}{n(n-1)(n-3)(n-7)(n-2)^2(n^5-15n^4+72n^3-156n^2+150n-42)},
\end{eqnarray}
\begin{eqnarray}\label{mu5}
\mu_{5}&=&\frac{\hat{\mu}_{{5}}{L}^{8}}{(n-3)(n-9)(n-2)^2 } (8\,{n}^{12}+26\,{n}^{11}-1489\,{n}^{10}+11130\,{n}^{9}-26362\,{n}^{8}-
75132\,{n}^{7}+705657\,{n}^{6}-2318456\,{n}^{5}\nonumber\\
&&\,\,\,\,\,\,\,\,\,\,\,\,\,\,\,\,\,\,\,\,\,\,\,\,\,\,\,\,\,\,\,\,\,\,\,\,\,\,\,\,\,\,\,\,\,\,\,\,\,\,\,\,\,\,\,\,\,\,\,\,\,\,+4461054\,{n}^{4}-
5484168\,{n}^{3}+4290516\,{n}^{2}-1968224\,n+405376)^{-1},
\end{eqnarray}
where $L$ is a scale factor that is related to the cosmological constant. Eq. \eqref{mu5} shows that $\mu_{5}$ is nonzero for all values of $n$, and so we can use quintic quasitopological gravity in all dimensions. This is the advantage of quintic quaitopological gravity in higher dimensions.\\
To obtain the field equations, we vary the action (\ref{Act2}) with respect to functions $f(r)$ and $g(r)$. They yield to the equations
\begin{equation}\label{equ1}
(1+2\hat{\mu}_{2} \Psi+3\hat{\mu}_{3}\Psi^2+4\hat{\mu}_{4}\Psi^3+5\hat{\mu}_{5}\Psi^4)N^{'}=0,
\end{equation}
\begin{equation}\label{equ2}
\bigg\{r^n\bigg(\hat{\mu}_{0}+\Psi+\hat{\mu}_{2} \Psi^2+\hat{\mu}_{3} \Psi^3+\hat{\mu}_{4}\Psi^4+\hat{\mu}_{5}\Psi^5\bigg)\bigg\}^{'}=0,
\end{equation}
respectively where we have used the substitution $f(r)=N^2(r)g(r)$. Equation (\ref{equ1}) shows that the function $N(r)$ should be a constant, and so we choose $N(r)=1$ which yields to $f(r)=g(r)$. Using $N=1$ in (\ref{equ2}), we reach to the equation
\begin{eqnarray}\label{equasli}
\hat{\mu}_{5} \Psi^5+\hat{\mu}_{4} \Psi^4+\hat{\mu}_{3} \Psi^3+\hat{\mu}_{2} \Psi^2+\Psi+\kappa=0,
\end{eqnarray}
where $\kappa$ is
\begin{eqnarray}
\kappa=\hat{\mu}_{0}-\frac{m}{r^{n}},
\end{eqnarray}
and $m$ is a constant of integration relating to the mass of the black hole. Choosing $\hat{\mu}_{5}=0$ in Eq. \eqref{equasli}, we get to a fourth order equation which can be solved analytically \cite{Deh1}. However, for $\hat{\mu}_{5}\neq0$, the fifth order equation (\ref{equasli}) can not be solved analytically and so numerical calculations are required. The obtained function $f(r)$ depends on parameters $r$, $n$, $m$, $k$, $L$, $\hat{\mu}_{5}$, $\hat{\mu}_{4}$, $\hat{\mu}_{3}$, $\hat{\mu}_{2}$ and $\hat{\mu}_{0}$.

\section{Thermodynamics of the solutions} \label{thermo}
Now, we intend to calculate the physical and thermodynamic properties of the solutions. Although we cannot find the solutions analytically, we are able to calculate the thermodynamic quantities exactly. The geometrical mass of this black hole is as follows
\begin{eqnarray}\label{geomass}
m=r_{+}^{n}\bigg(\hat{\mu}_{0}+k\frac{L^2}{r_{+}^2}+\hat{\mu}_{2} k^2\frac{L^4}{r_{+}^4}+\hat{\mu}_{3} k^3\frac{L^6}{r_{+}^6}+\hat{\mu}_{4} k^4\frac{L^8}{r_{+}^8}+\hat{\mu}_{5} k^5\frac{L^{10}}{r_{+}^{10}}\bigg),
\end{eqnarray}
where $r_{+}$ is defined as the radial coordinate of the outermost horizon of the black hole and is the positive root of the equation $f(r_{+})=0$. If we use the reference background 
\begin{eqnarray}\label{ADM1}
ds^2=-W^2(r)dt^2+\frac{dr^2}{V^2(r)}+r^2d\Omega^2,
\end{eqnarray}
and write the metric \eqref{metr} in this form, we get to a quasilocal mass by subtraction method. This mass depends on the choice of the reference background, so we use the limit $r\rightarrow\infty$ 
to obtain the mass of this black hole per unit volume $V_{n-1}$ as
\begin{eqnarray}
M=\frac{(n-1)m}{16 \pi L^2},
\end{eqnarray}
where $m$ is the mass parameter defined in Eq. \eqref{geomass}. To calculate the temperature of this black hole, we employ the analytic continuation of the metric. In this method, one uses the inverse of the period of
Euclidean time required for the absence of conical singularities in the Euclidean continuation of the geometry. Therefore, we first use the transformation $t\rightarrow i\tau$ in the Euclidean section of the metric. By involving the transformation $\tau\rightarrow \tau+\beta_{+}$ in this section, regularity is established at $r = r_{+}$, where $\beta_{+}$ is the inverse of the Hawking temperature. So, the temperature is gained at the event horizon $r_{+}$ as
\begin{eqnarray}\label{temperaure}
T&=&\bigg(\frac{f^{'}(r)}{4\pi}\bigg)_{r=r_{+}}\nonumber\\
&=&\frac {\left( n-10 \right){k}^{5}\hat{\mu}_{5}l^{10}+\left( n-8
 \right){k}^{4} \hat{\mu}_{4} l^{8}{r}^{2}+\left( n-6 \right){k}^{3}\hat{\mu}_{3} l^{6}{r}^{4}+\left(n-
 4\right){k}^{2}\hat{\mu}_{2} l^{4}{r}^{6}+\left( n-2 \right) k l^{2}  {r}^{8}+n \hat{\mu}_{0}{r
}^{10}}{4\pi l^2 r( {r}^{8}+2\,\hat{\mu}_{2}kl^2{r}^{6}+3\,\hat{\mu}_{3}{k}^{2}l^4{r}^{4}+4\,\hat{\mu}_{4}{k}^{3}{r}^{2}l^6+5\,\hat{\mu}_{5}{k}^{4}l^8)}.
\end{eqnarray}
In order to have an extreme black hole, the condition $T=0$ should be satisfied. For the obtained solutions, there is an extreme black hole for $k=0$ in flat specetime, while for $k=\pm 1$, there would be an extreme black hole with a horizon radius $r_{\rm ext}$ which is the largest real root of the equation
\begin{eqnarray}
\hat{\mu}_{5}{k}^{5} \left( n-10 \right) +\hat{\mu}_{4} {k}^{4} \left( n-8
 \right) {r_{\rm ext}}^{2}+\hat{\mu}_{3}{k}^{3} \left( n-6 \right) {r_{\rm ext}}^{4}+{k}^{2}\hat{\mu}_{2} \left( -
4+n \right) {r_{\rm ext}}^{6}+k\hat{\mu}_{0}  \left( n-2 \right) {r_{\rm ext}}^{8}+n{r_{\rm ext}
}^{10}=0.
\end{eqnarray}
We call the mass parameter of the extreme black hole $m_{\rm {ext}}$, which is $m_{\rm {ext}}=m(r_{+}=r_{\rm{ext}})$. \\ 
To calculate entropy of this black hole, we use the Wald formula \cite{Iyer} in which the entropy is given by  
\begin{eqnarray}\label{entropy}
s=-2\pi \oint d^{n-1} x \sqrt{\tilde{g}} \frac{\partial \mathcal L}{\partial R_{abcd}}\hat{\epsilon}_{ab}\hat{\epsilon}_{cd},
\end{eqnarray}
and the entropy density would be $S=s/V_{n-1}$. In Eq. \eqref{entropy}, $\tilde{g}$ is the determinant of the induced metric on the horizon and $\hat{\epsilon}_{ab}$ is the binormal of the horizon. Also, $Y=\frac{\partial \mathcal L}{\partial R_{abcd}}\hat{\epsilon}_{ab}\hat{\epsilon}_{cd}$, where $\mathcal{L}$ is the Lagrangian consists of all Einstein-Hilbert, Gauss-Bonnet, cubic, quartic and quintic quasitopological gravities. For the values of $Y$ in the cubic and quartic quasitopological gravities see Refs. \cite{Deh1,Myers3} respectively. To obtain the terms of the quintic form, we should vary all 24 terms in Lagrangian \eqref{quintic}. As these variated terms are too long, for economic reasons, we have mentioned just four terms as follow
\begin{eqnarray}
Y_{4}&=& 2R_{d}^{c}R_{e}^{d}R_{c}^{e}(R^{t}_{t}+R^{r}_{r})+3(R_{e}^{t}R_{t}^{e}+R_{e}^{r}R_{r}^{e})R_{b}^{a}R_{a}^{b},\nonumber\\
Y_{8}&=& R_{f}^{c}R_{cg}^{de}(R_{d}^{r}R_{re}^{fg}+R_{d}^{t}R_{te}^{fg})+R_{f}^{c}R_{ae}^{fg}(R_{r}^{a}R_{cg}^{re}+R_{t}^{a}R_{cg}^{te})+R_{b}^{a}R_{d}^{b}(R_{rg}^{de}R_{ae}^{rg}+R_{tg}^{de}R_{ae}^{tg})\nonumber\\
&&+R_{b}^{a}(R_{r}^{b}R_{f}^{r}R_{at}^{ft}+R_{t}^{b}R_{f}^{t}R_{ar}^{fr}-R_{r}^{b}R_{f}^{t}R_{at}^{fr}-R_{t}^{b}R_{f}^{r}R_{ar}^{ft})+R_{d}^{b}(R_{b}^{r}R_{r}^{c}R_{ct}^{dt}+R_{b}^{t}R_{t}^{c}R_{cr}^{dr}-R_{b}^{t}R_{r}^{c}R_{ct}^{dr}-R_{b}^{r}R_{t}^{c}R_{cr}^{dt}),\nonumber\\
Y_{23}&=& R_{fj}^{gh}R_{dh}^{ij}(R_{rg}^{rd}R_{ti}^{tf}
+R_{tg}^{td}R_{ri}^{rf}-R_{rg}^{td}R_{ti}^{rf}
-R_{tg}^{rd}R_{ri}^{tf})+R_{bi}^{ef} (R_{re}^{rb}R_{fj}^{th}R_{th}^{ij}+R_{te}^{tb}R_{fj}^{rh}R_{rh}^{ij}-R_{re}^{tb}R_{fj}^{rh}R_{th}^{ij}-
R_{te}^{rb}R_{fj}^{th}R_{rh}^{ij})\nonumber\\
&&+R_{ag}^{cd}(R_{cr}^{ar}R_{tj}^{gh}R_{dh}^{tj}+R_{ct}^{at}R_{rj}^{gh}R_{dh}^{rj}-R_{cr}^{at}R_{tj}^{gh} R_{dh}^{rj}-R_{ct}^{ar}R_{rj}^{gh}R_{dh}^{tj})+
R_{ce}^{ab}(R_{ar}^{cd}R_{bi}^{er}R_{dt}^{it}
+R_{at}^{cd}R_{bi}^{et}R_{dr}^{ir}-R_{ar}^{cd}
R_{bi}^{et}R_{dt}^{ir}\nonumber\\
&&-R_{at}^{cd}R_{bi}^{er}R_{dr}^{it})+R_{ce}^{ab}(R_{ag}^{cr}R_{br}^{ef}
R_{ft}^{gt}+R_{ag}^{ct}R_{bt}^{ef}R_{fr}^{gr}
-R_{ag}^{ct}R_{br}^{ef}R_{ft}^{gr}-R_{ag}^{cr}
R_{bt}^{ef}R_{fr}^{gt}), \nonumber\\
Y_{24}&=& (R_{fg}^{rd}R_{hi}^{tf}R_{rj}^{gh}R_{td}^{ij}
+R_{fg}^{td}R_{hi}^{rf}R_{tj}^{gh}R_{rd}^{ij}
-R_{fg}^{td}R_{hi}^{rf}R_{rj}^{gh}R_{td}^{ij}
-R_{fg}^{rd}R_{hi}^{tf}R_{tj}^{gh}R_{rd}^{ij})+
(R_{re}^{ab}R_{hi}^{er}R_{aj}^{th}R_{bt}^{ij}+
R_{te}^{ab}R_{hi}^{et}R_{aj}^{rh}R_{br}^{ij}\nonumber\\
&&-R_{re}^{ab}R_{hi}^{et}R_{aj}^{rh}R_{bt}^{ij}
-R_{te}^{ab}R_{hi}^{er}R_{aj}^{th}R_{br}^{ij})
+(R_{cr}^{ab}R_{tg}^{cd}R_{aj}^{gr}R_{bd}^{tj}+
R_{ct}^{ab}R_{rg}^{cd}R_{aj}^{gt}R_{bd}^{rj}
-R_{cr}^{ab}R_{tg}^{cd}R_{aj}^{gt}R_{bd}^{rj}
-R_{ct}^{ab}R_{rg}^{cd}R_{aj}^{gr}R_{bd}^{tj})+\nonumber\\
&&(R_{ce}^{rb}R_{fr}^{cd}R_{ti}^{ef}R_{bd}^{it}+
R_{ce}^{tb}R_{ft}^{cd}R_{ri}^{ef}R_{bd}^{ir}
-R_{ce}^{tb}R_{fr}^{cd}R_{ti}^{ef}R_{bd}^{ir}
-R_{ce}^{rb}R_{ft}^{cd}R_{ri}^{ef}R_{bd}^{it})+
(R_{ce}^{ar}R_{fg}^{ct}R_{hr}^{ef}R_{at}^{gh}+
R_{ce}^{at}R_{fg}^{cr}R_{ht}^{ef}R_{ar}^{gh}\nonumber\\
&&-R_{ce}^{at}R_{fg}^{cr}R_{hr}^{ef}R_{at}^{gh}
-R_{ce}^{ar}R_{fg}^{ct}R_{ht}^{ef}R_{ar}^{gh}).
\end{eqnarray}
If we add all 24 variated terms and do the integration in Eq. \eqref{entropy}, we get to 
\begin{eqnarray}
S_{5}= \frac{5(n-1)L^8k^4\hat{\mu}_{5}}{4(n-9)}r_{+}^{n-9}.
\end{eqnarray} 
Totalizing the obtained values of $S$ in all gravities, the total entropy density would be  
\begin{eqnarray}\label{enf}
S=\frac{r_{+}^{n-1}}{4}\bigg(1+2k\hat{\mu}_{2} \frac{(n-1)L^2}{(n-3)r_{+}^2}+3k^2\hat{\mu}_{3} \frac{(n-1)L^4}{(n-5)r_{+}^4}+4k^3\hat{\mu}_{4} \frac{(n-1)L^6}{(n-7)r_{+}^6}+5k^4\hat{\mu}_{5} \frac{(n-1)L^8}{(n-9)r_{+}^8}\bigg).
\end{eqnarray}
If we choose the values $\hat{\mu}_{2}=\hat{\mu}_{3}=\hat{\mu}_{4}=\hat{\mu}_{5}=0$ in Eq. \eqref{enf}, the entropy density reduces to the entropy of Einstein-Hilbert theory. So, in the presence of higher-order curvature theories, some corrections are added to Einstein's entropy.\\
To check the first law of thermodynamics, we first consider $S$ as an
extensive parameter for the mass $M(S)$, where $T$ is expressed as the intensive parameter with the definition
\begin{eqnarray}\label{firstlaw1}
T=\frac{\partial M}{\partial S}.
\end{eqnarray}
In order to obtain the quantity $T$ in the above equation, we use $\frac{\partial M}{\partial S}=\frac{\frac{\partial M}{\partial r_{+}}}{\frac{\partial S}{\partial r_{+}}}$. Comparing the calculated $T$ in Eq. \eqref{firstlaw1}, and the obtained temperature in Eq. \eqref{temperaure}, we come to an equality which confirms the first law of thermodynamics for the obtained solutions of this black hole as 
\begin{eqnarray}
dM=TdS.
\end{eqnarray}
\section{Physical structure of the solutions} \label{phys}
\begin{figure}
\center
\includegraphics[scale=0.5]{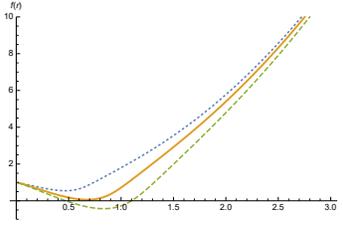}
\caption{\small{The asymptotically AdS solutions $f(r)$ versus $r$ in QQT gravity with $k=L=1$, $n=4$, $\hat{\mu}_{0}=1$, $\hat{\mu}_{2}=0.2$, $\hat{\mu}_{3}=0.1$, $\hat{\mu}_{4}=0.06$, $\hat{\mu}_{5}=0.01$ and $m<m_{\rm ext}$, $m=m_{\rm ext}$ and $m>m_{\rm ext}$ from up to down, respectively.}
\label{fig1}}
\end{figure}

\begin{figure}
\center
\includegraphics[scale=0.5]{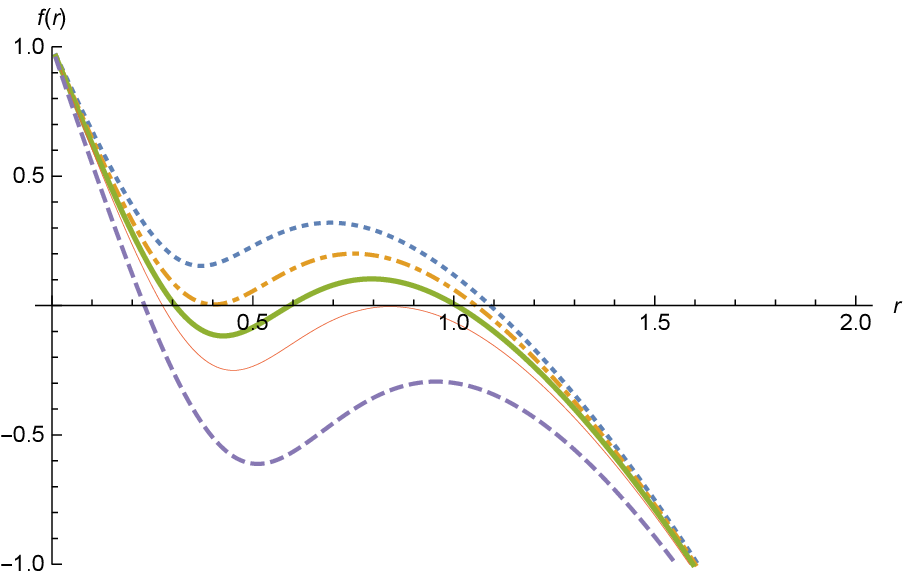}

\caption{\small{The asymptotically dS solutions $f(r)$ versus $r$ in QQT gravity with $k=L=1$, $n=5$, $\hat{\mu}_{0}=-1$, $\hat{\mu}_{2}=0.4$, $\hat{\mu}_{3}=0.01$, $\hat{\mu}_{4}=0.001$ and $\hat{\mu}_{5}=0.0005$ and $m<m_{\rm ext}$, $m=m_{\rm ext}$, $m_{\rm ext}<m<m_{\rm cri}$, $m=m_{\rm cri}$ and $m>m_{\rm cri}$ from up to down, respectively.} \label{fig2}}
\end{figure}

\begin{figure}
\center
\includegraphics[scale=0.5]{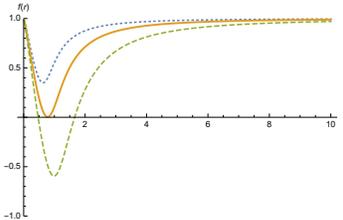}
\caption{\small{The asymptotically flat solution $f(r)$ versus $r$ in QQT gravity with $k=L=1$, $n=4$, $\hat{\mu}_{0}=0$, $\hat{\mu}_{2}=0.2$, $\hat{\mu}_{3}=0.1$, $\hat{\mu}_{4}=0.06$ and $\hat{\mu}_{5}=0.01$ and $m<m_{\rm ext}$, $m=m_{\rm ext}$ and $m>m_{\rm ext}$ from up to down, respectively.} \label{fig3}}
\end{figure}

Now, we want to investigate the physical structure of the solutions. For this purpose, we have plotted $f(r)$ versus $r$ in Anti de Sitter (AdS), de Sitter (dS) and flat spacetimes in Figs. \eqref{fig1}, \eqref{fig2} and \eqref{fig3} respectively. In this figures, there are two values $m_{\rm {ext}}$ and $m_{\rm {cri}}$ which are the mass parameter for the smaller and larger roots
of $T=0$ respectively. We should emphasis that $m_{\rm {ext}}=m(r_{+}=r_{\rm ext})$ and $m_{\rm {cri}}=m(r_{+}=r_{\rm cri})$. It is clear in all figures that the function $f(r)$ has a finite value at the origin, and as $r\rightarrow\infty$, it goes to $+\infty$, $-\infty$ and a constant value, for $\rm AdS$, $\rm dS$ and flat spacetimes respectively. In Fig. \eqref{fig1}, for the given values of the parameters, there is an extreme black hole with $m_{\rm {ext}}=1.3$ which is shown by the solid brown diagram. For $m<m_{\rm {ext}}$, there is a naked singularity and for $m>m_{\rm {cri}}$, there is a black hole with two inner and outer horizons. Therefore, for large values of mass parameter $m$, the possibility of having a black hole with two horizons is more.\\
In Fig. \ref{fig2}, there are two extreme black holes with masses $m_{\rm {ext}}=0.3$ and $m_{\rm {cri}}=0.53$ in dS spacetime, which are shown by dashed-dotted brown and thin solid pink diagrams respectively. It is worth mentioning that the value of $r_{\rm ext}$ is smaller than that of $r_{\rm cri}$. For $m<m_{\rm {ext}}$, there is a nonextreme black hole with one horizon $r_{+\rm max}$, where $r_{+\rm max}>r_{\rm {cri}}$. Also for $m_{\rm {ext}}<m<m_{\rm {cri}}$, there is a black hole with three horizons and for $m>m_{\rm {cri}}$, there is a nonextreme black hole with horizon at $r_{+\rm min}$, where $r_{+\rm min}<r_{\rm {ext}}$. 
In Fig. \eqref{fig3}, we have repeated this investigation for the asymptotically flat solution $f(r)$ with different values of $m$. It should be noted that for constant values of all parameters except $m$, there is an $m_{\rm {ext}}$ for which there is an extreme black hole. For $m<m_{\rm {ext}}$ and $m>m_{\rm {ext}}$, there is a naked singularity and a black hole with two horizons respectively.\\

\section{Thermal stability}\label{stability}
\begin{figure}
\centering
\subfigure[Temperature $T$]{\includegraphics[scale=0.27]{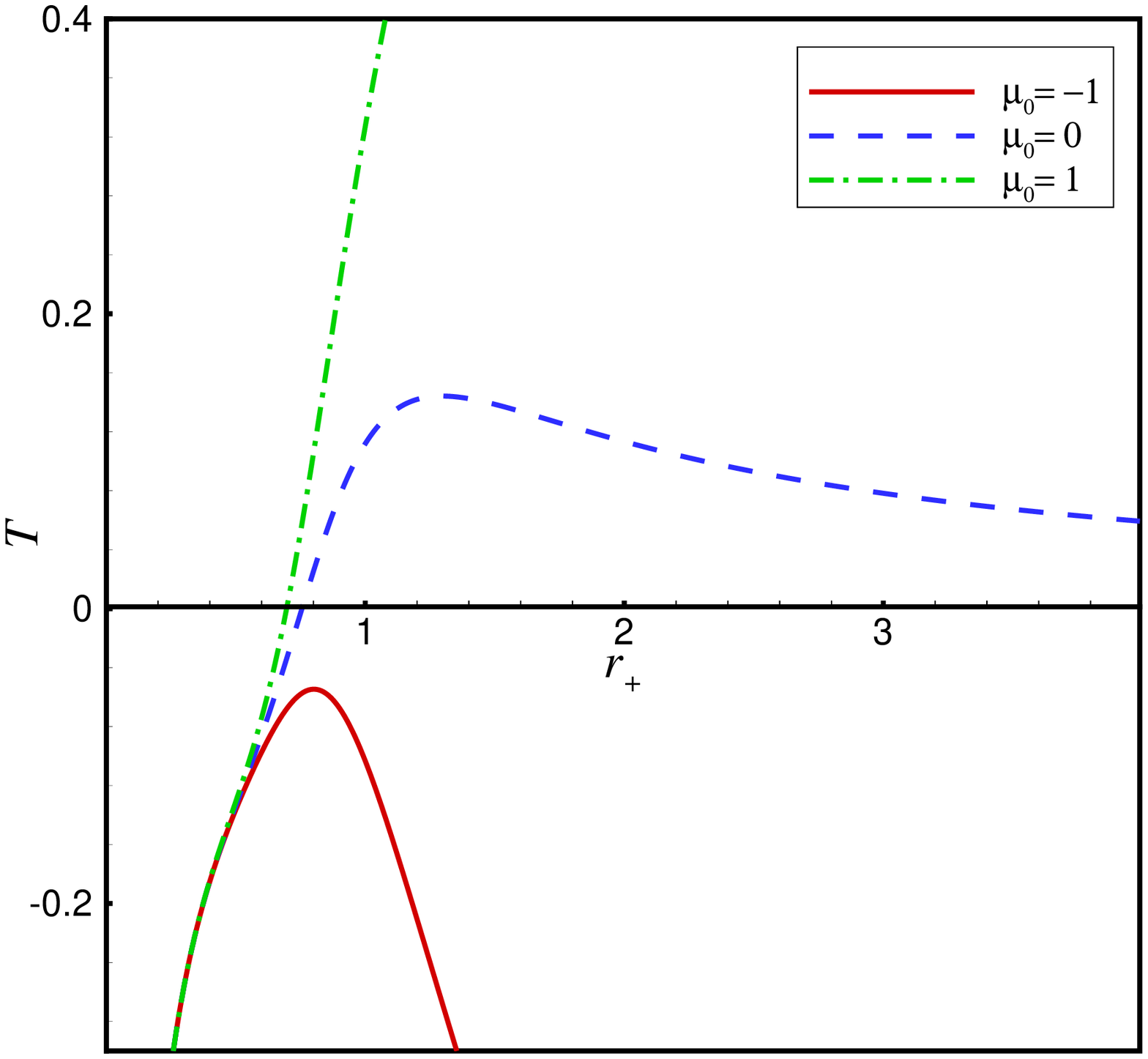}\label{fig4a}}\hspace*{.2cm}
\subfigure[Heat capacity $C_{Q}$]{\includegraphics[scale=0.27]{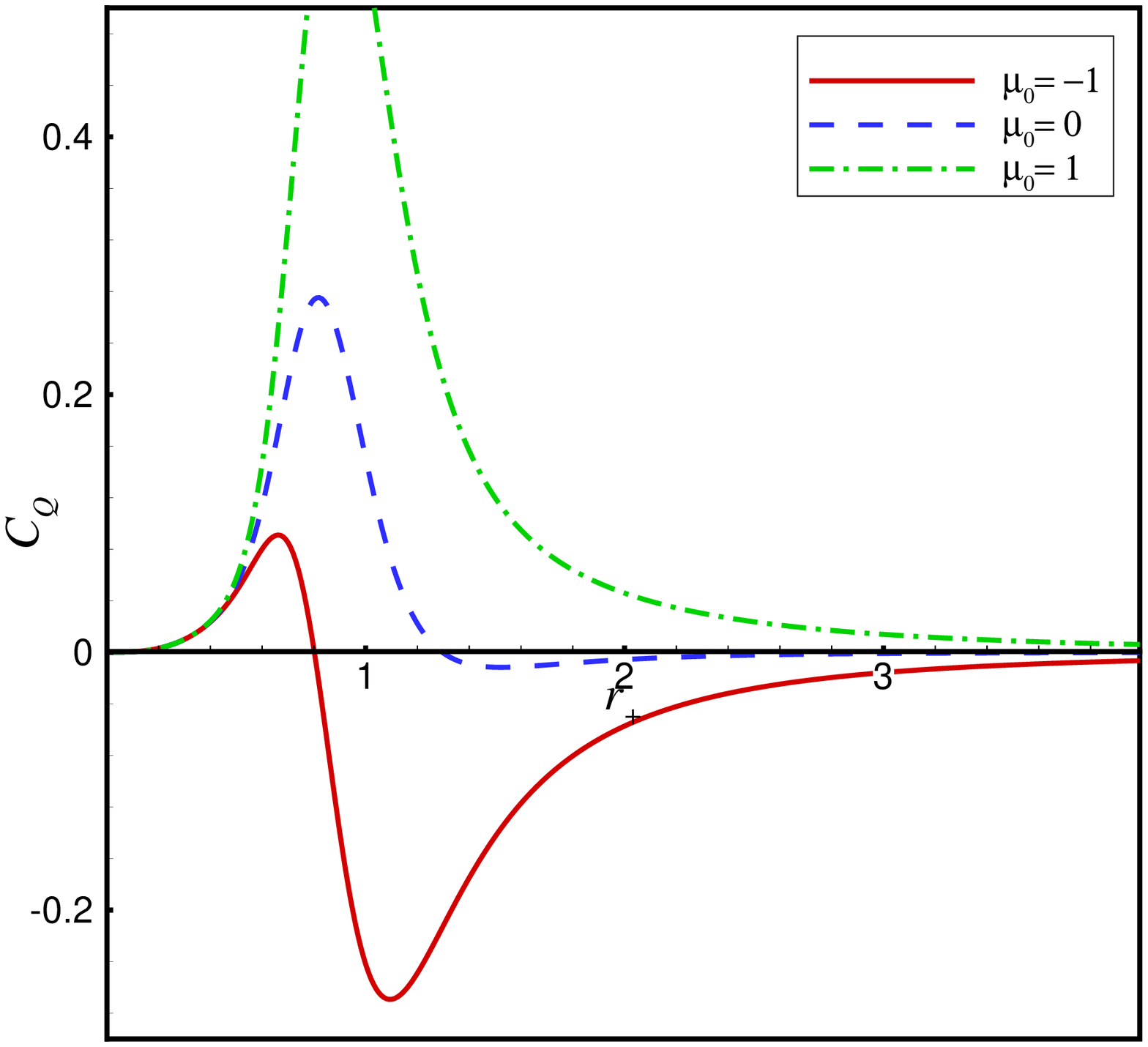}\label{fig4b}}\caption{$T$ and $C_{Q}$ versus $r_{+}$ for different $\hat{\mu}_{0}$ with $k=L=1$, $n=5$, $\hat{\mu}_{5}=0.04$, $\hat{\mu}_{4}=0.06$, $\hat{\mu}_{3}=0.09$ and $\hat{\mu}_{2}=0.07$.}\label{figure4}
\end{figure}

\begin{figure}
\centering
\subfigure[Temperature $T$]{\includegraphics[scale=0.27]{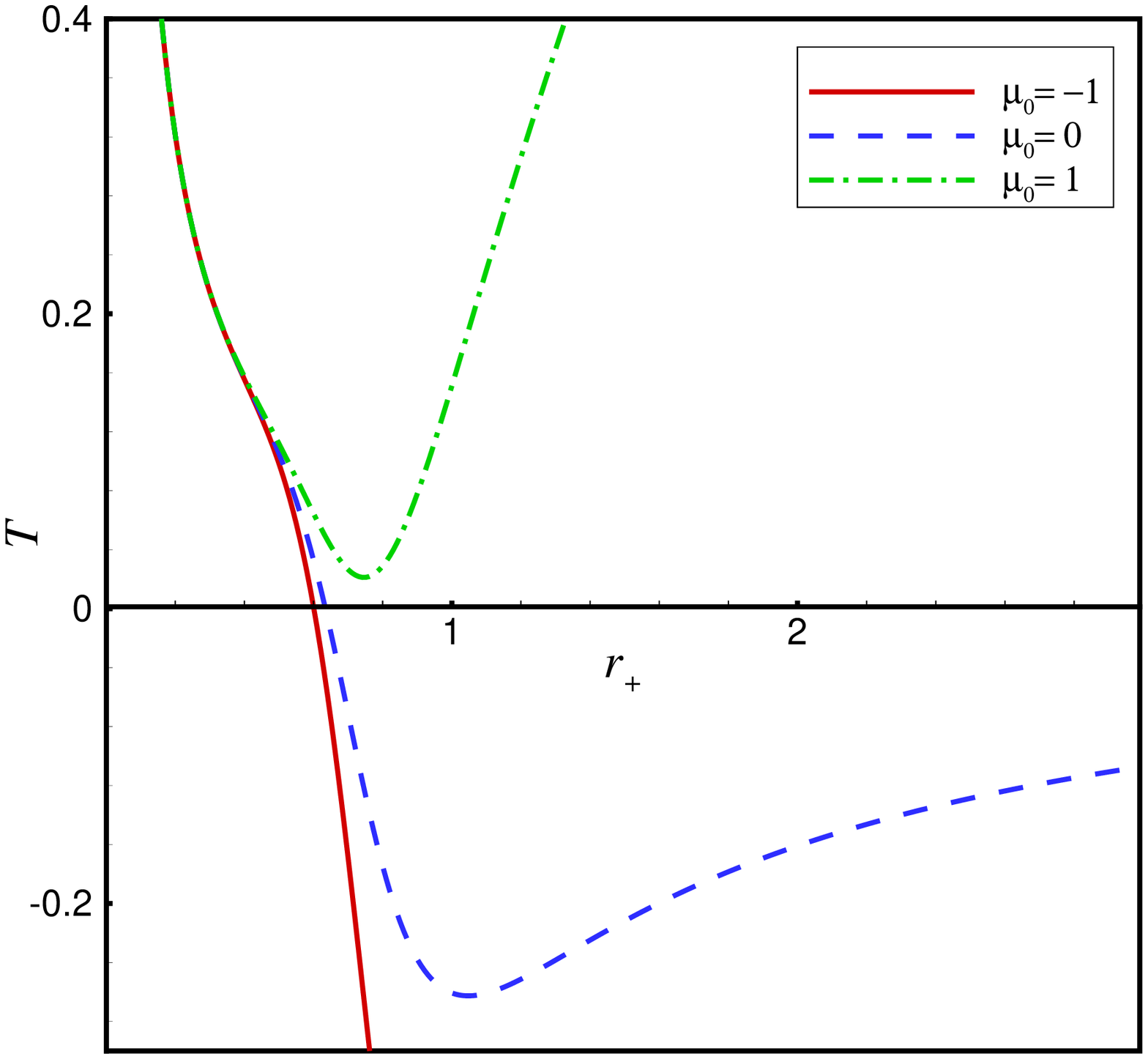}\label{fig5a}}\hspace*{.2cm}
\subfigure[Heat capacity $C_{Q}$]{\includegraphics[scale=0.27]{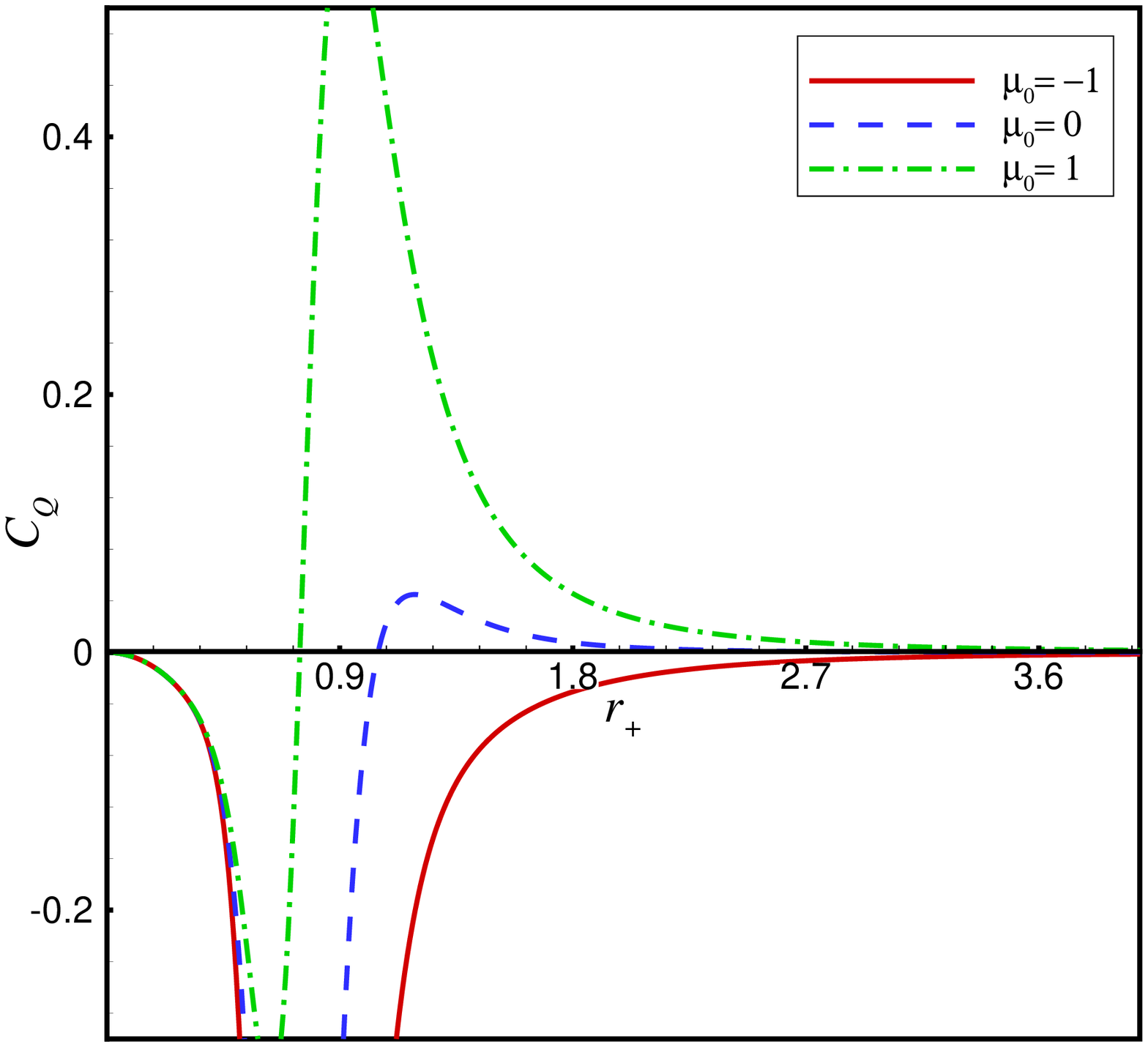}\label{fig5b}}\caption{$T$ and $C_{Q}$ versus $r_{+}$ for different $\hat{\mu}_{0}$ with $L=1$, $k=-1$, $n=6$, $\hat{\mu}_{5}=0.03$, $\hat{\mu}_{4}=0.03$, $\hat{\mu}_{3}=0.09$ and $\hat{\mu}_{2}=0.07$.}\label{figure5}
\end{figure}

Now, we would like to check out the thermal stability of higher dimensional quintic quasitopological black hole solutions. To study this, we can analyze the behavior of its energy $M(S)$ with respect to small variations of the $S$. For the local stability, $M(S)$ should be a convex function of $S$. We probe thermal stability of this black hole in canonical ensemble. In this ensemble, the black hole is thermally stable if the heat capacity $C_{Q}$ and temperature $T$, both have positive values. The heat capacity is defined as
\begin{equation}\label{heat}
C_{Q}=\frac{\partial ^2 M}{\partial S^2}=\frac{r^{-n+10}}{(n-1)\pi L^2 r^2}\times\frac{(AB-r^2 CD)}{ A^3},
\end{equation}
where 
\begin{eqnarray}
A&=& 5\,{k}^{4}\hat{\mu}_{5}\,{L}^{8}+4\,{k}^{3}\hat{\mu}_{4}\,{L}^{6}{r}^{2}+3\,{k}^{2}\hat{\mu}_{3}
\,{L}^{4}{r}^{4}+2\,k\hat{\mu}_{2}\,{L}^{2}{r}^{6}+{r}^{8},\nonumber\\
B&=&- \left(n-10 \right) {k}^{5}\hat{\mu}_{5} {L}^{10}+ \left( n-8
 \right) {k}^{4}\hat{\mu}_{4}\,{L}^{8}{r}^{2}+3\, \left( {\it n}-6 \right) {k}
^{3}\hat{\mu}_{3}\,{L}^{6}{r}^{4}+5\, \left( {\it n}-4 \right) {k}^{2}\hat{\mu}_{2}\,{L
}^{4}{r}^{6}+7\, \left( {\it n}-2 \right) k{L}^{2}{r}^{8}+9\,{\it n}\hat{\mu}_{0}\,{r}^{10},\nonumber\\
C&=&\left( {\it n}-10 \right) {k}^{5}\hat{\mu}_{5}\,{L}^{10}+ \left( {\it n}-8 \right) {k}^{4}\hat{\mu}_{4}\,{L}^{8}{r}^{2}+ \left( {\it n}-6 \right) {k}^{3
}\hat{\mu}_{3}\,{L}^{6}{r}^{4}+ \left( {\it n}-4 \right) {k}^{2}\hat{\mu}_{2}\,{L}^{4}{
r}^{6}+ \left( {\it n}-2 \right) k{L}^{2}{r}^{8}+{\it n}\,\hat{\mu}_{0}\,{r}^{10},\nonumber\\
D&=& 8\,{L}^{6}{k}^{3}\hat{\mu}_{4}+12\,{L}^{4}{k}^{2}\hat{\mu}_{3}\,{r}^{2}+12\,{L}^{2}k
\hat{\mu}_{2}\,{r}^{4}+8\,{r}^{6}.
\end{eqnarray}
If we substitute $k=0$ in Eqs. \eqref{temperaure} and \eqref{heat}, we get to the simple forms of equations 
\begin{eqnarray}\label{simple}
C_{Q}=\frac{n \hat{\mu}_{0}r^{2-n}}{(n-1)\pi L^2}\,\,\,\,,\,\,\,\,T=\frac{n\hat{\mu}_{0}r}{4\pi L^2},
\end{eqnarray} 
where indicate that the positivity of these quantities depends on the positivity of $\hat{\mu}_{0}$. This concludes that for $k=0$, higher dimensional quintic quasitopological black hole is thermally stable in just AdS spacetime (or for the value $\hat{\mu}_{0}>0$). However, for $k=\pm 1$, as we can not get to simple forms of equations for $C_{Q}$ and $T$, we have plotted them in Figs. \eqref{figure4} and \eqref{figure5} to probe the stability of the black hole. We have plotted the behavior of both $T$ and $C_{Q}$ for $k=1$ and $k=-1$ in Figs. \eqref{figure4} and \eqref{figure5} respectively. In Fig. \eqref{figure4}, the dS solutions are not thermally stable because $T$ is negative for all values of $r_{+}$. For AdS solutions, $C_{Q}$ is positive for all $r_{+}$ and therefore, stability depends on $T$. In this spacetime, there is a $r_{+ \rm minAdS1}$, where $T>0$ for $r_{+}>r_{+ \rm minAdS1}$. For flat spacetime, there are two $r_{+ \rm minflat}$ and $r_{+ \rm maxflat}$ that $T>0$ for $r_{+}>r_{+ \rm minflat}$ and $C_{Q}>0$ for $r_{+}<r_{+ \rm maxflat}$. Therefore, flat solutions are thermally stable for the intersection of these two regions which is $r_{+ \rm minflat}<r_{+}<r_{+ \rm maxflat}$. \\
In Fig. \eqref{figure5}, we have shown the stability of the black hole for $k=-1$ in AdS, dS and flat spacetimes. It is clear that $C_{Q}$ is negative for all values of $r_{+}$ in dS solutions which yields to instability. For AdS solutions, $T$ is positive and so we go to $C_{Q}$ to recognize stability. We see that there is an $r_{+ \rm minAdS2}$, where $C_{Q}$ is positive for $r_{+}>r_{+ \rm minAdS2}$. For flat spacetime, $T$ and $C_{Q}$ are positive for $r_{+}\lesssim 0.6$ and $r_{+} \gtrsim 0.98$ respectively. These two regions have no intersection and therefore the stability is unachievable for flat solutions with $k=-1$ for given parameters in the caption.

\section{CONCLUDING REMARKS}\label{result}
Recently, quintic quasitopological gravity in five dimensions has been proposed which its structures on asymptotically AdS spacetimes might be duals for a broad class of CFTs. As some important theories like string and M-theories predict the dimensions of the spacetime to be more than four, so we completed this process and constructed a theory of quintic quasitopological gravity in higher dimensions $n\geq 5$. In spherically symmetric spacetimes, this gravity yields to the field equation with at most second order derivatives.\\
Then, we obtained the field equations of this gravity in $(n+1)$-dimensional static spacetime. In this gravity, we led to a fifth-order equation which could not be solved exactly and so we calculated the function $f(r)$ numerically and we probed the physical behavior of $f(r)$ in different spacetimes. For AdS and flat spacetimes, there is an extreme black hole with $m=m_{\rm ext}$, that for $m<m_{\rm ext}$ there is a naked singularity, but for $m>m_{\rm ext}$, there is a black hole with two horizons. For dS solutions, there are two extreme black holes with mass parameters $m_{\rm ext}$ and $m_{\rm cri}$. For $m<m_{\rm ext}$ and $m>m_{\rm cri}$, there are nonextreme black holes with one horizon and for $m_{\rm ext}<m<m_{\rm cri}$, there is a black hole with three horizons. \\
We also obtained the thermodynamic quantities for this black hole. We calculated the mass per unit volume $V_{n-1}$ by the subtraction method, entropy density by Wald formula and temperature by analytic continuation and then checked the first law of thermodynamics. \\
Then we probed the thermal stability of this black hole. For $k=0$, this black hole has thermal stability just for AdS solutions. For $k=\pm 1$, we plotted two figures to probe thermal stability.\\
For the chosen parameters with $k=1$, this black hole is not stable for dS solutions. There are also three parameters $r_{+\rm{minAdS1}}$, $r_{+\rm{minflat}}$ and $r_{+\rm{mamxflat}}$, where black hole is thermally stable in AdS and flat spacetimes for $r_{+}>r_{+\rm{minAdS1}}$ and $r_{+\rm{minflat}}<r_{+}<r_{+\rm{maxflat}}$ respectively.\\
For $k=-1$ and the given parameters, there is a $r_{+\rm{minAdS2}}$, where the black hole is stable just in AdS spacetime for $r_{+}>r_{+\rm{minAdS2}}$. So, for the selected parameters, AdS solutions have thermal stability compared to dS ones for each values of $k$, but flat solutions are stable with just $k=1$.\\
This higher dimensional quintic quasitopological gravity can provide more backgrounds for study. In future, we tend to promote this gravity in various spacetimes and investigate different kinds of solutions.

\acknowledgments{AB, FN and ARO would like to thank Jahrom University.  ARO is grateful to Institute for Research in Fundamental Sciences (IPM) for their warm hospitality, where a part of this work was completed.  MG would like to thank Payam Noor University.}
\subsection{APPENDIX}\label{app}
The coefficients $a_{i}$'s, $b_{i}$'s and $c_{i}$'s are defined as:
\begin{eqnarray}
a_{1}&=& 1,\,\,\,\,\,a_{2}=\frac{3(3n-5)}{8(2n-1)(n-3)},\,\,\,\,\,a_{3}=-\frac{3(n-1)}{(2n-1)(n-3)},\,\,\,\,\,a_{4}=\frac{3(n+1)}{(2n-1)(n-3)},\nonumber\\
&& a_{5}=\frac{6(n-1)}{(2n-1)(n-3)},\,\,\,\,\,
a_{6}=-\frac{3(3n-1)}{2(2n-1)(n-3)},\,\,\,\,\,a_{7}=\frac{3(n+1)}{8(2n-1)(n-3)}
\end{eqnarray}
\begin{eqnarray}
&&b_{1}=-(n-1)(n^7-3n^6-29n^5+170n^4-349n^3+348n^2-180n+36)\nonumber\\
&&b_{2}=-4(n-3)(2n^6-20n^5+65n^4-81n^3+13n^2+45n-18)\nonumber\\
&&b_{3}=-64(n-1)(3n^2-8n+3)(n^2-3n+3)\nonumber\\
&&b_{4}=-(n^8-6n^7+12n^6-22n^5+114n^4-345n^3+468n^2-270n+54)\nonumber\\
&&b_{5}=16(n-1)(10n^4-51n^3+93n^2-72n+18)\nonumber\\
&&b_{6}=-32(n-1)^2(n-3)^2(3n^2-8n+3)\nonumber\\
&&b_{7}=64(n-2)(n-1)^2(4n^3-18n^2+27n-9)\nonumber\\
&&b_{8}=-96(n-1)(n-2)(2n^4-7n^3+4n^2+6n-3)\nonumber\\
&&b_{9}=16(n-1)^3(2n^4-26n^3+93n^2-117n+36)\nonumber\\
&&b_{10}=n^5-31n^4+168n^3-360n^2+330n-90\nonumber\\
&&b_{11}=2(6n^6-67n^5+311n^4-742n^3+936n^2-576n+126)\nonumber\\
&&b_{12}=8(7n^5-47n^4+121n^3-141n^2+63n-9)\nonumber\\
&&b_{13}=16n(n-1)(n-2)(n-3)(3n^2-8n+3)\nonumber\\
&&b_{14}=8(n-1)(n^7-4n^6-15n^5+122n^4-287n^3+297n^2-126n+18).
\end{eqnarray}
\begin{eqnarray}
c_{1}&=&\frac{1}{n-2}( 22\,{n}^{12}+98\,{n}^{11}-4227\,{n}^{10}+26488\,{n}^{9}-34298\,{n}^{8}-314764\,{n}^{7}+1879963\,{n}^{6}-5179230\,{n}^{5}+8667296\,{n}^{4}\nonumber\\
&&\,\,\,\,\,\,\,\,\,\,\,\,\,\,\,\,\,-
9278000\,{n}^{3}+6209228\,{n}^{2}-2352032\,n+379200)\nonumber\\
c_{2}&=& 9\,{n}^{11}+34\,{n}^{10}-1541\,{n}^{9}+11499\,{n}^{8}-25758
\,{n}^{7}-81964\,{n}^{6}+660233\,{n}^{5}-1886059\,{n}^{4}+3046869\,{n}
^{3}-2977682\,{n}^{2}\nonumber\\
&&+1666312\,n-41192\nonumber\\
c_{3}&=&\frac{1}{2(n-2)}( -58\,{n}^{12}-162\,{n}^{11}+10663\,{n}^{10}-84812\,{
n}^{9}+229322\,{n}^{8}+436556\,{n}^{7}-5176607\,{n}^{6}+18005330\,{n}^
{5}\nonumber\\
&&\,\,\,\,\,\,\,\,\,\,\,\,\,\,\,\,\,\,\,\,\,\,\,\,\,\,-35943244\,{n}^{4}+45563680\,{n}^{3}-36695932\,{n}^{2}+17330208\,n-
3674560)\nonumber\\
c_{4}&=&-\frac{2(n-1)}{n-2}( 9\,{n}^{11}+34\,{n}^{10}-1541\,{n}^{9}+11499\,{n}^{8}-25758\,{n}^{7}-81964\,{n}^{6}+660233
\,{n}^{5}-1886059\,{n}^{4}+3046869\,{n}^{3}\nonumber\\
&&\,\,\,\,\,\,\,\,\,\,\,\,\,\,\,\,\,\,\,\,\,\,\,\,\,\,\,\,\,-2977682\,{n}^{2}+1666312\,n-411920)\nonumber\\ 
c_{5}&=&\frac{1}{4(n-2)}(208\,{n}^{13}-4737\,{n}^{12}+40968\,{n}^{11}-159932\,{n}^{10}+101251\,{n}^{9}+1850607\,{n}^{8}-9772230\,{n}^{7}+27253898
\,{n}^{6}\nonumber\\
&&\,\,\,\,\,\,\,\,\,\,\,\,\,\,\,\,\,\,\,\,\,\,\,\,\,-50334197\,{n}^{5}+65342916\,{n}^{4}-60349728\,{n}^{3}+
38913248\,{n}^{2}-16207200\,n+3316864)\nonumber\\
c_{6}&=&\frac{1}{n-2}(-296\,{n}^{13}+5380\,{n}^{12}-47491\,{n}^{11}+235224\,{n}^{10}-501416\,{n}^{9}-1195535\,{n}^{8}+12548311\,{n}^{7}-45635482
\,{n}^{6}\nonumber\\
&&\,\,\,\,\,\,\,\,\,\,\,\,\,\,\,\,\,\,+100350946\,{n}^{5}-146329207\,{n}^{4}+143219210\,{n}^{3}-
91132732\,{n}^{2}+34380784\,n-5893664)\nonumber\\
c_{7}&=&\frac{1}{4(n-2)}(184\,{n}^{14}-3251\,{n}^{13}+28056\,{n}^{12}-109604\,{n}^{11}+6501\,{n}^{10}+1605461\,{n}^{9}-5747494\,{n}^{8}+3380818\,{
n}^{7}\nonumber\\
&&\,\,\,\,\,\,\,\,\,\,\,\,\,\,\,\,\,\,\,\,\,\,\,\,+36068661\,{n}^{6}-144617644\,{n}^{5}+293846956\,{n}^{4}-
373665156\,{n}^{3}+300154032\,{n}^{2}-139373056\,n\nonumber\\
&&\,\,\,\,\,\,\,\,\,\,\,\,\,\,\,\,\,\,\,\,\,\,\,\,\,+28457216)\nonumber\\
c_{8}&=&\frac{1}{4(n-2)}(596\,{n}^{13}-7977\,{n}^{12}+71966\,{n}^{11}-505962\,{n}^{10}+2089493\,{n}^{9}-2365377\,{n}^{8}-20061508\,{n}^{7}+
119539756\,{n}^{6}\nonumber\\
&&\,\,\,\,\,\,\,\,\,\,\,\,\,\,\,\,\,\,\,\,\,\,\,\,\,-335696499\,{n}^{5}+581268584\,{n}^{4}-651267392\,{n
}^{3}+461836336\,{n}^{2}-188752160\,n+33860352)\nonumber\\
c_{9}&=&\frac{1}{16(n-2)}(-184\,{n}^{15}+3155\,{n}^{14}-31372\,{n}^{13}+214234\,{n}^{12}-1011489\,{n}^{11}+2804783\,{n}^{10}+374252\,{n}^{9}-44192768\,{n}^{8}\nonumber\\
&&\,\,\,\,\,\,\,\,\,\,\,\,\,\,\,\,\,\,\,\,\,\,\,\,\,\,\,\,+224431715\,{n}^{7}-655954220\,{n}^{6}+1293485398\,{n
}^{5}-1792474880\,{n}^{4}+1739485312\,{n}^{3}-1131595440\,{n}^{2}\nonumber\\
&&\,\,\,\,\,\,\,\,\,\,\,\,\,\,\,\,\,\,\,\,\,\,\,\,\,\,\,\,+
442875968\,n-78459392)\nonumber\\
c_{10}&=&\frac{1}{2(n-2)}(304\,{n}^{14}-5487\,{n}^{13}+51364\,{n}^{12}-296956\,{n}^{11}+1020583\,{n}^{10}-1134859\,{n}^{9}-8135394\,{n}^{8}+
52879112\,{n}^{7}\nonumber\\
&&\,\,\,\,\,\,\,\,\,\,\,\,\,\,\,\,\,\,\,\,\,\,\,\,\,-168012561\,{n}^{6}+347472004\,{n}^{5}-498259688\,{n}
^{4}+497441450\,{n}^{3}-331820224\,{n}^{2}+132631584\,n\nonumber\\
&&\,\,\,\,\,\,\,\,\,\,\,\,\,\,\,\,\,\,\,\,\,\,\,\,\,\,\,\,-23851392)\nonumber
\end{eqnarray}
\begin{eqnarray}
c_{11}&=&\frac{1}{4(n-2)}(-244\,{n}^{13}+4709\,{n}^{12}-34468\,{n}^{11}+95172\,{n}^{10}+152097\,{n}^{9}-1923839\,{n}^{8}+6353794\,{n}^{7}-11131154
\,{n}^{6}\nonumber\\
&&\,\,\,\,\,\,\,\,\,\,\,\,\,\,\,\,\,\,\,\,\,\,\,\,\,\,+10232149\,{n}^{5}-1781288\,{n}^{4}-6422656\,{n}^{3}+6551632
\,{n}^{2}-2066336\,n-21504)\nonumber
\end{eqnarray}
\begin{eqnarray}
c_{12}&=&\frac{1}{8(n-2)}(416\,{n}^{14}-10647\,{n}^{13}+86586\,{n}^{12}-223848\,{n}^{11}-764407\,{n}^{10}+6904499\,{n}^{9}-18735836\,{n}^{8}+
11482750\,{n}^{7}\nonumber\\
&&\,\,\,\,\,\,\,\,\,\,\,\,\,\,\,\,\,\,\,\,\,\,\,\,\,+69049061\,{n}^{6}-239246282\,{n}^{5}+400589060\,{n}^
{4}-410760584\,{n}^{3}+261506352\,{n}^{2}-94377920\,n\nonumber\\
&&\,\,\,\,\,\,\,\,\,\,\,\,\,\,\,\,\,\,\,\,\,\,\,\,\,+14506368)\nonumber
\end{eqnarray}
\begin{eqnarray}
c_{13}&=&\frac{1}{16(n-2)}(-184\,{n}^{15}+4003\,{n}^{14}-34770\,{n}^{13}+206558\,{n}^{12}-1209685\,{n}^{11}+6001605\,{n}^{10}-16647870\,{n}^{9}
-3841080\,{n}^{8}\nonumber\\
&&\,\,\,\,\,\,\,\,\,\,\,\,\,\,\,\,\,\,\,\,\,\,\,\,\,\,\,\,\,+218943659\,{n}^{7}-902806270\,{n}^{6}+2083343490\,{n
}^{5}-3136302944\,{n}^{4}+3171015856\,{n}^{3}-2094129968\,{n}^{2}\nonumber\\
&&\,\,\,\,\,\,\,\,\,\,\,\,\,\,\,\,\,\,\,\,\,\,\,\,\,\,\,\,\,+
819673024\,n-144243200)\nonumber
\end{eqnarray}
\begin{eqnarray}
c_{14}&=&\frac{1}{2(n-2)}(388\,{n}^{14}-4716\,{n}^{13}+29243\,{n}^{12}-136746\,{n}^{11}+450540\,{n}^{10}-132929\,{n}^{9}-8134503\,{n}^{8}+48332850
\,{n}^{7}\nonumber\\
&&\,\,\,\,\,\,\,\,\,\,\,\,\,\,\,\,\,\,\,\,\,\,\,\,\,-155977854\,{n}^{6}+329810835\,{n}^{5}-478872342\,{n}^{4}+
476176930\,{n}^{3}-310557920\,{n}^{2}+119520320\,n\nonumber\\
&&\,\,\,\,\,\,\,\,\,\,\,\,\,\,\,\,\,\,\,\,\,\,\,\,\,\,\,\,-20516736)\nonumber
\end{eqnarray}
\begin{eqnarray}
c_{15}&=&\frac{1}{8(n-2)}(664\,{n}^{14}-9139\,{n}^{13}+57128\,{n}^{12}-185108\,{n}^{11}+159381\,{n}^{10}+1223517\,{n}^{9}-6100234\,{n}^{8}+14465638
\,{n}^{7}\nonumber\\
&&\,\,\,\,\,\,\,\,\,\,\,\,\,\,\,\,\,\,\,\,\,\,\,\,\,-20079091\,{n}^{6}+16176660\,{n}^{5}-10250920\,{n}^{4}+
14943120\,{n}^{3}-20032144\,{n}^{2}+11852416\,n-2285952)\nonumber
\end{eqnarray}
\begin{eqnarray}
c_{16}&=&\frac{1}{2}(540\,{n}^{13}-10293\,{n}^{12}+81315\,{n}^{11}-291890\,{n}^{10}+61415\,
{n}^{9}+3807202\,{n}^{8}-16976001\,{n}^{7}+38237858\,{n}^{6}\nonumber\\
&&\,\,\,\,\,\,-48573651
\,{n}^{5}+26466351\,{n}^{4}+15817758\,{n}^{3}-37469132\,{n}^{2}+
25084592\,n-6246112)\nonumber
\end{eqnarray}
\begin{eqnarray}
c_{17}&=&\frac{1}{16(n-2)}(432\,{n}^{15}-4127\,{n}^{14}+19469\,{n}^{13}-170554\,{n}^{12}+1675605\,{n}^{11}-9507738\,{n}^{10}+29711901\,{n}^{9}\nonumber\\
&&\,\,\,\,\,\,\,\,\,\,\,\,\,\,\,\,\,\,\,\,\,\,\,\,\,\,-
40384306\,{n}^{8}-54483045\,{n}^{7}+383370077\,{n}^{6}-912069066\,{n}^
{5}+1331612328\,{n}^{4}-1311989568\,{n}^{3}\nonumber\\
&&\,\,\,\,\,\,\,\,\,\,\,\,\,\,\,\,\,\,\,\,\,\,\,\,\,\,+870354912\,{n}^{2}-
355966464\,n+67907584)\nonumber\\
c_{18}&=&\frac{1}{2(n-2)}(62\,{n}^{15}-261\,{n}^{14}+82\,{n}^{13}-34985\,{n}^{12}+465930\,{n}^{11}-2557591\,{n}^{10}+6958394\,{n}^{9}-5370935\,{n}^{8}\nonumber\\
&&\,\,\,\,\,\,\,\,\,\,\,\,\,\,\,\,\,\,\,\,\,\,\,\,\,-27811996\,{n}^{7}+116102040\,{n}^{6}-231876220\,{n}^{5}+291631996
\,{n}^{4}-242759516\,{n}^{3}+131704680\,{n}^{2}\nonumber\\
&&\,\,\,\,\,\,\,\,\,\,\,\,\,\,\,\,\,\,\,\,\,\,\,\,\,-43058976\,n+6606272)\nonumber
\end{eqnarray}
\begin{eqnarray}
c_{19}&=&\frac{1}{4(n-2)}(-656\,{n}^{15}+8832\,{n}^{14}-54341\,{n}^{13}+172912\,{n}^{12}-46160\,{n}^{11}-2326941\,{n}^{10}+11290819\,{n}^{9}-
23788482\,{n}^{8}\nonumber\\
&&\,\,\,\,\,\,\,\,\,\,\,\,\,\,\,\,\,\,\,\,\,\,\,\,\,\,+3603422\,{n}^{7}+115748491\,{n}^{6}-336672132\,{n}^{
5}+513849092\,{n}^{4}-485503192\,{n}^{3}+287262016\,{n}^{2}\nonumber\\
&&\,\,\,\,\,\,\,\,\,\,\,\,\,\,\,\,\,\,\,\,\,\,\,\,\,\,\,\,-98692160\,
n+15146752)\nonumber
\end{eqnarray}
\begin{eqnarray}
c_{20}&=&\frac{1}{8(n-2)}(-1328\,{n}^{15}+24603\,{n}^{14}-201582\,{n}^{13}+847816\,{n}^{12}-1334949\,{n}^{11}-4313683\,{n}^{10}+32443416\,{n}^{9}\nonumber\\
&&\,\,\,\,\,\,\,\,\,\,\,\,\,\,\,\,\,\,\,\,\,\,\,\,\,-106895622\,{n}^{8}+248652911\,{n}^{7}-456679514\,{n}^{6}+663949044\,{
n}^{5}-732715856\,{n}^{4}+587021544\,{n}^{3}\nonumber\\&&\,\,\,\,\,\,\,\,\,\,\,\,\,\,\,\,\,\,\,\,\,\,\,\,\,-326614080\,{n}^{2}+
116003840\,n-20182272)\nonumber
\end{eqnarray}
\begin{eqnarray}
c_{21}&=&\frac{1}{8(n-2)}(184\,{n}^{16}-3339\,{n}^{15}+27760\,{n}^{14}-131212\,{n}^{13}+355561\,{n}^{12}-357883\,{n}^{11}-1572146\,{n}^{10}+
11274022\,{n}^{9}\nonumber\\
&&\,\,\,\,\,\,\,\,\,\,\,\,\,\,\,\,\,\,\,\,\,\,\,\,\,\,-44976839\,{n}^{8}+128367156\,{n}^{7}-256851408\,{n}^
{6}+335947624\,{n}^{5}-240332008\,{n}^{4}+11641104\,{n}^{3}\nonumber\\&&\,\,\,\,\,\,\,\,\,\,\,\,\,\,\,\,\,\,\,\,\,\,\,\,\,+134199392
\,{n}^{2}-104249344\,n+26702336)\nonumber
\end{eqnarray}
\begin{eqnarray}
c_{22}&=&\frac{1}{2(n-2)}(-284\,{n}^{15}+4973\,{n}^{14}-37942\,{n}^{13}+144773\,{n}^{12}-109479\,{n}^{11}-1875825\,{n}^{10}+12234317\,{n}^{9}-
45166705\,{n}^{8}\nonumber\\
&&\,\,\,\,\,\,\,\,\,\,\,\,\,\,\,\,\,\,\,\,\,\,\,\,\,\,+119677671\,{n}^{7}-240530864\,{n}^{6}+367236029\,{n}
^{5}-416310288\,{n}^{4}+337180200\,{n}^{3}-183807888\,{n}^{2}\nonumber\\&&\,\,\,\,\,\,\,\,\,\,\,\,\,\,\,\,\,\,\,\,\,\,\,\,\,\,+60487840
\,n-9120896)\nonumber
\end{eqnarray}
\begin{eqnarray}
c_{23}&=&\frac{1}{4(n-2)}(-8\,{n}^{15}+2019\,{n}^{14}-38926\,{n}^{13}+337600\,{n}^{12}-1605181\,{n}^{11}+3785705\,{n}^{10}+1659444\,{n}^{9}-
45775086\,{n}^{8}\nonumber\\
&&\,\,\,\,\,\,\,\,\,\,\,\,\,\,\,\,\,\,\,\,\,\,\,\,\,\,+176674471\,{n}^{7}-400284742\,{n}^{6}+611871600\,{n}
^{5}-648934536\,{n}^{4}+469233704\,{n}^{3}-218274992\,{n}^{2}\nonumber\\
&&\,\,\,\,\,\,\,\,\,\,\,\,\,\,\,\,\,\,\,\,\,\,\,\,\,\,+57719936\,n-6343424)\nonumber
\end{eqnarray}
\begin{eqnarray}
c_{24}&=&\frac{1}{4}(184\,{n}^{15}-3179\,{n}^{14}+25777\,{n}^{13}-115454\,{n}^{12}+228481\,
{n}^{11}+522238\,{n}^{10}-5783003\,{n}^{9}+23848974\,{n}^{8}\nonumber\\
&&\,\,\,\,\,\,\,-64717433
\,{n}^{7}+128477225\,{n}^{6}-193789406\,{n}^{5}+224224860\,{n}^{4}-
195140632\,{n}^{3}+120313912\,{n}^{2}\nonumber\\
&&\,\,\,\,\,\,\,\,-46440128\,n+8345216)
\end{eqnarray}


\end{document}